\documentclass[english,prl,twocolumn,superscriptaddress]{revtex4}
\usepackage[T1]{fontenc}
\usepackage[latin9]{inputenc}
\usepackage{color}
\usepackage{amsmath}
\usepackage{graphicx}

\makeatletter
\@ifundefined{textcolor}{}
{%
 \definecolor{BLACK}{gray}{0}
 \definecolor{WHITE}{gray}{1}
 \definecolor{RED}{rgb}{1,0,0}
 \definecolor{GREEN}{rgb}{0,1,0}
 \definecolor{BLUE}{rgb}{0,0,1}
 \definecolor{CYAN}{cmyk}{1,0,0,0}
 \definecolor{MAGENTA}{cmyk}{0,1,0,0}
 \definecolor{YELLOW}{cmyk}{0,0,1,0}
 }

\makeatother

\usepackage{babel}

\begin{document}

\title{Testing for Multipartite Quantum Nonlocality Using Functional Bell
Inequalities }

\author{Q. Y. He}

\affiliation{Centre for Quantum-Atom Optics, Swinburne University of Technology,
Melbourne, Australia}

\author{E. G. Cavalcanti}

\affiliation{Centre for Quantum Dynamics, Griffith University, Brisbane QLD 4111,
Australia}

\author{M. D. Reid}

\affiliation{Centre for Quantum-Atom Optics, Swinburne University of Technology,
Melbourne, Australia}

\author{P. D. Drummond}

\affiliation{Centre for Quantum-Atom Optics, Swinburne University of Technology,
Melbourne, Australia}
\begin{abstract}
\textcolor{black}{\normalsize We show that arbitrary functions of
continuous variables, e.g. position and momentum, can be used to generate
tests that distinguish quantum theory from local hidden variable theories.
By optimising these functions, we obtain more robust violations of
local causality than obtained previously. We analytically calculate
the optimal function and include the effect of nonideal detectors
and noise, revealing that optimized functional inequalities are resistant
to standard forms of decoherence. These inequalities could allow a
loophole-free Bell test with efficient homodyne detection.}{\normalsize \par}
\end{abstract}
\maketitle
Bell famously showed that the predictions of quantum mechanics (QM)
are not always compatible with local hidden variable theories (LHV)
\cite{Bell}. Surprisingly, this fundamental result, which underpins
the field of quantum information, has not been rigorously tested \cite{moreloop}.
There are no experiments yet that can eliminate all LHV, either due
to low detection efficiencies \cite{loophole,recent prl crit eff}
or lack of causal separation. Rigorous tests are also needed to fully
implement some quantum information protocols, like that of Ekert \cite{Ekert}
which employs a Bell inequality (BI) as a test of security in a cryptographic
scheme. All of these early tests and protocols employed quantum measurements
with discrete outcomes of spin or particle number.

In this Letter, we develop \emph{functional moment} inequalities to
test for quantum nonlocality. We then use variational calculus to
optimize the choice of measured function. As a result, we obtain Bell
nonlocality for larger losses and for greater degrees of decoherence
than possible previously. The outcome can be feasibly tested in the
laboratory, since the detectors required are efficient quadrature
detectors. More generally, functional nonlocality measures could lead
to new applications in quantum information. The important advantage
is a much greater robustness to noise and loss.

As well as potentially overcoming the loophole problem mentioned above,
formalisms to test LHV for continuous variables provide an opportunity
for testing QM in new environments, and give a better understanding
of the origin of the nonlocal features of QM. This is particularly
true given that entanglement \cite{peres} alone does not guarantee
failure of LHV for mixed states \cite{werner}.

With this objective, there is the fundamental question of how to \emph{quantify}
the strength of nonlocality, in the absence of a single test for nonlocality
that is necessary and sufficient for any quantum state. Mermin \cite{mermin-1}
used as a measure the deviation of the QM prediction from the LHV
bound, based on a particular BI. A second strategy discussed recently
by Cabello \emph{et al}. \cite{recent prl crit eff} is to quantify
the strength of nonlocality by the robustness of the violation with
respect to a decoherence parameter. In this approach one determines
the critical efficiency $\eta$ or the critical degree of purity $p$
required for a violation. Here, we evaluate all three measures to
show strong correlations between them. 

Recently, Cavalcanti \emph{et al.} (CFRD) showed \cite{caval} that
Bell inequalities can be derived for the case of observables with
continuous and unbounded outcomes, like position and momentum. This
approach is significant in establishing that quantum nonlocality does
not rely on the discreteness of the measurement outcomes. Continuous
variable (cv) inequalities also provide an avenue to understanding
how manifestations of quantum nonlocality can be manipulated by choice
of observable.

The original CFRD inequality \cite{caval} is $|\langle\prod_{k=1}^{N}(x_{k}+ip_{k})\rangle|^{2}\leq\langle\Pi_{k=1}^{N}(x_{k}^{2}+p_{k}^{2})\rangle,$
where $x_{k}$, $p_{k}$ are the outcomes of two arbitrary measurements,
represented in QM by observables $\hat{X_{k}}$, $\hat{P_{k}}$, at
site $k$ \cite{LHS}. Where $\hat{X}$ and $\hat{P}$ are quadrature
measurements with canonical position and momentum commutation relations,
CFRD showed that the symmetric state $\bigl\{|0\rangle^{\otimes N/2}|1\rangle^{\otimes N/2}+|1\rangle^{\otimes N/2}|0\rangle^{\otimes N/2}\bigl\}/\sqrt{2}$
violates the inequality for $N\geq10$. In this case, the states $|0\rangle$,
$|1\rangle$ are eigenstates of $a^{\dagger}a$ where $\hat{a}=\hat{X}+i\hat{P}$,
so the prediction could in principle be tested with photonic Greenberger-Horne-Zeilinger
(GHZ) states produced in the laboratory \cite{expghz}. Note that
in the above state there are $N$ field modes but only $N/2$ photons.
It can be prepared from a $N/2$-photon GHZ state $\{|H\rangle^{\otimes N/2}+|V\rangle^{\otimes N/2}\}/\sqrt{2},$
where $|H\rangle,\ |V\rangle$ represent horizontally or vertically
polarized single-photon states, by passing each photon through a polarizing
beam splitter. These violations are robust with loss. The critical
efficiency $\eta_{crit}$ required for violation tends to $\eta_{crit}\rightarrow0.81$,
as $N\rightarrow\infty$. Quadrature measurements with local oscillators
are highly efficient, with reported efficiencies of $99\%$. However,
generation losses from mode-matching can degrade the experimental
efficiency, so $81\%$ is still a challenging practical benchmark.

Instead, we introduce a\emph{ functional moment} Bell inequality by
considering arbitrary functions of the outcomes at each site. This
new approach to nonlocality utilizes a general functional optimization
of continuous variable observables. We~\textcolor{black}{find the
optimal function that maximizes a violation of the inequality for
a given efficiency $\eta$ and state purity $p$. We show that the
optimal function has the form }\textcolor{black}{\emph{$x/(1+\varepsilon_{N}x^{2})$,$\ $}}\textcolor{black}{{}
where $\varepsilon_{N}$ is a parameter related to $N$ and $\eta$.
This gives an inequality which is violated by the GHZ states of (\ref{eq:GHZ})
for $N\geq5$. The violation increases exponentially with $N$, while
$\eta_{crit}$ decreases asymptotically to $0.69$ for a pure state
(with $p=1$), thus} dramatically reducing both the number of modes
required, and the required efficiency. 

When the functions correspond to a simple binning of a cv observable
to give binary outcomes \cite{cvbin}, our inequalities reduce to
those of Mermin \cite{mermin-1}. We extend the analysis of Mermin
and Acin \emph{et al}. \cite{Acin }, and calculate results for homodyne
detection for more feasible types of state. We find that $(\eta p^{2}){}_{crit}=2^{(1-2N)/N}\pi$,
which gives a critical efficiency for a pure state at large $N$ of
$\eta=0.79$\textcolor{black}{.}

\textit{\textcolor{black}{Functional Moment Inequalities.$\ $}}\textbf{\textit{\textcolor{black}{{}
}}}We present a proof of the functional moment inequality taking explicit
account of functions of measurements that can be made at each of $N$
spatially separated sites. We denote the measurement made on the system
at the $k$-th site by $X_{k}^{\theta}$, and the outcome of the measurement
by $x_{k}^{\theta}$, where $\theta$ represents a choice of measurement
parameter. Bell's assumption that LHV can describe the outcomes implies
that the measurable moments $\langle x_{1}^{\theta}x_{2}^{\phi}\ldots x_{N}^{\varphi}\rangle$
can be expressed in terms of a set of hidden variables $\lambda$
as\begin{equation}
\langle x_{1}^{\theta}x_{2}^{\phi}\ldots x_{N}^{\varphi}\rangle=\int_{\lambda}d\lambda P(\lambda)\langle x_{1}^{\theta}\rangle_{\lambda}\langle x_{2}^{\phi}\rangle_{\lambda}\ldots\langle x_{N}^{\varphi}\rangle_{\lambda}\ ,\label{eq:LHV_correlations}\end{equation}
where $\langle x_{k}^{\theta}\rangle_{\lambda}$ is the average of
$x_{k}^{\theta}$ given a LHV state $\lambda$. Next we construct,
for each site $k$, real functions of two observables $f_{k}(x_{k}^{\theta})$,
$g_{k}(x_{k}^{\theta'})$, and define the complex function: $F_{k}=f_{k}(x_{k}^{\theta})+ig_{k}(x_{k}^{\theta'})$.
The complex moment $\langle F_{1}F_{2}...F_{N}\rangle$ can be expressed
in terms of real-valued expressions of the type $\langle f_{1}(x_{1}^{\theta})g_{2}(x_{2}^{\phi'})...f_{N}(x_{N}^{\varphi})\rangle,$
etc. Of course, $f_{k}(x_{k}^{\theta})$ is an observable obtained
from $x_{k}^{\theta}$ by local post-measurement processing. Eq. \eqref{eq:LHV_correlations}
must therefore also be valid for $\langle f(x_{1}^{\theta})f(x_{2}^{\phi})\ldots f(x_{N}^{\varphi})\rangle=\int_{\lambda}d\lambda P(\lambda)\langle f(x_{1}^{\theta})\rangle_{\lambda}\langle f(x_{2}^{\phi})\rangle_{\lambda}\ldots\langle f(x_{N}^{\varphi})\rangle_{\lambda}.$
For an LHV, the expectation value of products of the $F_{k}$ must
satisfy:\begin{equation}
\langle F_{1}\ldots F_{N}\rangle=\int_{\lambda}d\lambda P(\lambda)\langle F_{1}\rangle_{\lambda}\ldots\langle F_{N}\rangle_{\lambda}\ ,\label{eq:F_k_moments}\end{equation}
where $\langle F_{k}\rangle_{\lambda}\equiv\langle f_{k}(x_{k}^{\theta})\rangle_{\lambda}+i\langle g_{k}(x_{k}^{\theta'})\rangle_{\lambda}.$
From \eqref{eq:F_k_moments}, the following inequality must therefore
hold:\begin{eqnarray}
|\langle F_{1}F_{2}...F_{N}\rangle|^{2} & \leq & \int d\lambda P(\lambda)|\langle F_{1}\rangle_{\lambda}|^{2}...|\langle F_{N}\rangle_{\lambda}|^{2}\ .\label{eq:F_k_mod_squared}\end{eqnarray}
Now for any particular value of $\lambda,$ the statistics predicted
for $f_{k}(x_{k})$ must have a non-negative variance, i.e., $\langle f_{k}(x_{k})\rangle_{\lambda}^{2}\leq\langle f_{k}(x_{k})^{2}\rangle_{\lambda}.$
Writing \eqref{eq:F_k_mod_squared} explicitly in terms of the $f_{k}$'s
and using this variance inequality we arrive at the CFRD inequality
with functional moments:\begin{equation}
\left|\left\langle \prod_{k=1}^{N}[f_{k}(x_{k}^{\theta})+ig_{k}(x_{k}^{\theta'})]\right\rangle \right|^{2}\leq\left\langle \prod_{k=1}^{N}[f_{k}(x_{k}^{\theta})^{2}+g_{k}(x_{k}^{\theta'})^{2}]\right\rangle .\label{eq:moment cv BI}\end{equation}

\textcolor{black}{We will measure the violation of this inequality
by the ratio of the left- ($LHS$) and right-hand sides ($RHS$).
Defining the Bell observable $B=LHS/RHS$, failure of LHV is demonstrated
when $B>1$. In order to get stronger violation of local causality,
we optimize the function of observables by considering \begin{equation}
\frac{\delta B}{\delta f_{k}(g_{k})}=0\ .\label{eq:optimal function}\end{equation}
Here, we consider the class of entangled states \begin{equation}
|\psi\rangle=(|0\rangle^{\otimes r}|1\rangle^{\otimes(N-r)}+|1\rangle^{\otimes r}|0\rangle^{\otimes(N-r)})/\sqrt{2}\ .\label{eq:GHZ}\end{equation}
Thus $r=N$ corresponds to extreme photon-number-correlated states},
\textcolor{black}{a superposition of a state with $0$ photons at
all sites and a state with $1$ photon at each site. Next, we consider
how to optimize the function $f_{k}$ and $g_{k}$ to generate a robustly
violated inequality, including losses and noise.}

\textcolor{black}{We use variational calculus to find the optimal
function using the condition of Eq. (\ref{eq:optimal function}).
For simplicity, we assume the functions $f_{k}$ and $g_{k}$ are
odd. The $LHS$ can be maximized by choosing orthogonal angles, while
the $RHS$ is invariant with angles. We find that}

\begin{equation}
B_{N}=\frac{2^{N-1}(\frac{2}{\pi})^{\frac{N}{2}}(\prod_{k=1}^{N}I_{k}^{+}+\prod_{k=1}^{N}I_{k}^{-})^{2}}{\prod_{k=1}^{r}I_{k}\prod_{k=r+1}^{N}I_{k}^{0}+\prod_{k=1}^{r}I_{k}^{0}\prod_{k=r+1}^{N}I_{k}}\ ,\label{eq:ideal symmetric}\end{equation}
\textcolor{black}{where $I_{k}^{\pm}=2\int e^{-2x^{2}}xf_{k}^{\pm}dx$,
$I_{k}=4\int x^{2}e^{-2x^{2}}[(f_{k}^{+})^{2}+(f_{k}^{-})^{2})]dx$,
and $I_{k}^{0}=\int e^{-2x^{2}}[(f_{k}^{+})^{2}+(f_{k}^{-})^{2})]dx$
are different integrals for $x$ which contribute to the expectation
values in both sides of inequality (\ref{eq:moment cv BI}). Here
$f_{k}^{\pm}=f_{k}\pm g_{k}$, and the factor }$e^{-2x^{2}}$\textcolor{black}{{}
~was obtained from the joint probability of observables. Requiring
$\delta B_{N}/\delta f_{k}^{\pm}=0$$ $, we find the optimal condition:
$f_{k}(x)=\pm g_{k}(x)$. The components of complex functions $f_{k}$,
$g_{k}$ are the same at each site, and have the form\begin{equation}
f_{k}(x)=g_{k}(x)=\frac{x}{1+\varepsilon_{N}x^{2}}\ .\label{eq:optimum_function}\end{equation}
 For the even $N$ case, it is optimal to choose $r=N/2$. Then $\varepsilon_{N}$
is independent of $N$, but has to be calculated numerically since
it satisfies a nonlinear integral equation: $\varepsilon_{N}=4I^{0}/I$. }

\begin{figure}[h]
\textcolor{black}{\includegraphics[width=0.7\columnwidth]{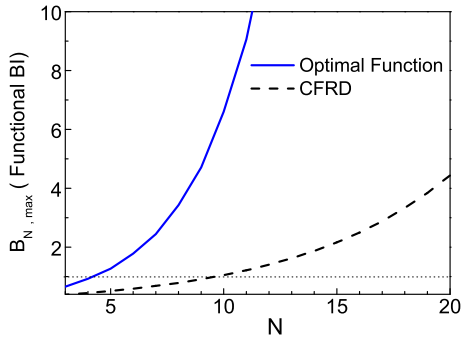}}

\textcolor{black}{\caption{\textcolor{black}{Maximum violations of functional cv inequality with
GHZ states} as a function of the number of modes.\textcolor{black}{{}
}~The violations using~\textcolor{black}{{} the optimal function (solid)}
are much stronger than the CFRD result (dashed). \label{fig:ideal Bell value}}
}
\end{figure}

\textcolor{black}{For $N$ an odd number, }the greatest violations
occur for $r=(N-1)/2$.~\textcolor{black}{{} The optimal function has
the same form as in (\ref{eq:optimum_function}) except that the parameter
$\varepsilon_{N}$ changes to $\varepsilon'_{N}$, where: \begin{equation}
\varepsilon'{}_{N}\equiv\varepsilon_{N}\left[\frac{N\varepsilon_{N}^{+}-\varepsilon_{N}^{-}}{N\varepsilon_{N}^{+}+\varepsilon_{N}^{-}}\right]\ ,\label{eq:odd_optimize}\end{equation}
and $\varepsilon_{N}^{\pm}=\varepsilon_{N}\pm4$. However, the numerical
value of $\varepsilon_{N}$ and $\varepsilon'_{N}$ now depend on
$N$, as the integral equation (\ref{eq:odd_optimize}) for odd values
of $N$ is $N$-dependent. This provides better violations of \eqref{eq:moment cv BI}
than any other arbitrary function, provided $N\geq5$. The maximum
$B_{N}$ value with this optimal choice is shown in Fig. \ref{fig:ideal Bell value},
compared with the CFRD result which uses a simple correlation function. }

\textcolor{black}{}%

\textit{\textcolor{black}{Binned cv outcomes for Mermin-Klyshko inequality
}}\textcolor{black}{(MK)~}\textit{\textcolor{black}{{} }}\textcolor{black}{\cite{other mer Klyshko }}\textit{\textcolor{black}{.$\ $}}\textcolor{black}{{}
}We will also briefly consider binning methods. Specifically, we define
the binning functions $f_{k}(x)=g_{k}(x)=f_{bin}(x)=+1$ i\textcolor{black}{f
$x\geq0$ and $-1$ otherwise. For such discrete outcomes, the original
formalisms of Mermin and Klyshko \cite{mermin-1,other mer Klyshko }
can be used. The CFRD inequality for discrete outcomes reduces to
that of Mermin \cite{mermin-1}, as can be seen by noting that $[f_{bin}]^{2}=1$.
Here it is known that the Bell inequality introduced for this discrete
case by Klyshko is stronger. Defining $F_{k}=f_{bin}(x_{k}^{\theta})+if_{bin}(x_{k}^{\theta'})$
and $\Pi_{N}=\prod_{k=1}^{N}F_{k}$, we can use the MK inequality
$|S_{N}|\leq1\ ,$ where $S_{N}=2^{-N/2}[\mathrm{Re}\{\Pi_{N}\}\pm\mathrm{Im\{}\Pi_{N}\}]$,
for $N$ even, and $S_{N}=2^{-(N-1)/2}\mathrm{Re}(\mathrm{Im})\{\Pi_{N}\}$
or $ $$S_{N}=2^{-(N-1)/2}\sqrt{(\mathrm{Re}\{\Pi_{N}\})^{2}+(\mathrm{Im}\{\Pi_{N}\})^{2}}$
for $N$ odd. These BI have been considered recently for the case
of extreme photon-number correlated states, where $r=N$, by Acin
}\textcolor{black}{\emph{et al}}\textcolor{black}{. We can also define
$\Pi_{N}$ by exchanging the local observables, to obtain similar
inequalities.}

\textcolor{black}{We generalize this approach to account for more
general angles and states. }We find an optimal violation of $B_{N}=\frac{\sqrt{2}}{2}(\frac{4}{\pi})^{N/2}$,
for arbitrary $r$ and $N$, with \textcolor{black}{the optimal phases:
$\theta_{k}=(-1)^{N+1}\pi(k-1)/(2N)$, }$\theta_{k}^{'}=\theta_{k}+\pi/2$
for $k\leq r$, and $\theta_{k}=(-1)^{N}\pi(k-1)/(2N)$, $\theta_{k}^{'}=\theta_{k}-\pi/2$
for $k>r$. This result has been presented by Acin \textit{et al.
}\cite{Acin } for the special case of $r=N$. We confirm the exponential
increase with number of sites $N$, but also make the observation
that the violation occurs for all types of states of the form (\ref{eq:GHZ}),
independently of $r$. This contrasts with the result for the CRFD
inequality, which requires $r\sim N/2$ for violation. However, as
explained earlier, the states with $r=N/2$ are straightforwardly
feasible given a polarisation GHZ state, as opposed to the extreme
photon-number correlated states considered by Acin \emph{et al}. Therefore,
this is a very important experimental advantage. Violation of the
MK inequality with binning is possible for $N\geq3$, but, as we will
see, this strategy is sensitive to losses and noise.

\textit{\textcolor{black}{Sensitivity to loss and state impurity.
}}The value of the Bell observable $B_{N}$ increases with the number
of sites $N$, so this is suggestive of a strategy that will allow
genuine loophole-free violations of local causality. However, it may
be argued that since increasing the number of sites will increase
the number of detectors required, there will be no advantage. Only
careful calculation of the Bell observable $B_{N}$ including the
detection efficiency $\eta$ can determine whether the strategy is
advantageous. 

Loss is modeled as follows. The field modes $a_{k}$ at each site
are independently coupled to a second mode $a_{k,vac}$ respectively,
assumed to be in a vacuum. Photons are lost from the field into the
vacuum mode, the strength of coupling determining the rate of loss.
This beam splitter model gives the final detected and vacuum mode
in terms of the inputs $a$ and $a_{vac}$ \begin{eqnarray}
a_{out} & = & \sqrt{\eta}a+\sqrt{1-\eta}a_{vac}\ ,\nonumber \\
a_{vac,out} & = & \sqrt{1-\eta}a-\sqrt{\eta}a_{vac}\ ,\end{eqnarray}
where $\eta$ is the efficiency, the probability of detecting a photon
after coupling. Since we only measure the {}``$a_{out}$'' not the
{}``$a_{vac,out}$'', we need to trace over the latter modes to
obtain the final density operator $\rho_{out}$ for the detected modes
after loss. We can also examine the effect of impurity, by considering
a state $\rho'=p|\psi\rangle\langle\psi|+(1-p)\rho_{mix}$, where
$\rho_{mix}$ is the mixed state obtained with a model for decoherence
in the occupation-number basis, i.e. $\rho_{mix}=[|0\rangle^{\otimes r}|1\rangle^{\otimes(N-r)}\langle0|^{\otimes r}\langle1|^{\otimes(N-r)}+|1\rangle^{\otimes r}|0\rangle^{\otimes(N-r)}\langle1|^{\otimes r}\langle0|^{\otimes(N-r)}]/2$,
and $p$ is the probability the system is in the original pure state
\eqref{eq:GHZ}\textcolor{black}{.$\ $ }

Including the effect of detection inefficiencies and noise, the parameter
$\varepsilon_{N}$ is changed to $\varepsilon_{N}(\eta)$ for the
optimum function. For $N$ even we find that\begin{eqnarray}
\varepsilon_{N}(\eta) & = & \frac{2\eta\varepsilon_{N}}{2\eta+(1-\eta)\varepsilon_{N}}\ ,\nonumber \\
B_{N} & = & 2^{N-2}\left[\frac{2(I^{+})^{4}(\eta p)^{2}}{\pi I^{o}C}\right]^{\frac{N}{2}}\ ,\label{eq:Even Bell}\end{eqnarray}
where $\varepsilon_{N}$ is defined as before, and $C=\eta I+(1-\eta)I^{0}$.
For the case of odd $N$ the relevant integral equations change, giving\textcolor{black}{{}
a modified (and slightly reduced) Bell variable $B'_{N}$, where:} 

\begin{eqnarray}
\varepsilon'_{N}(\eta) & = & \varepsilon_{N}(\eta)\frac{N\varepsilon_{N}^{+}(\eta)-\varepsilon_{N}(\eta)\varepsilon_{N}^{-}/\varepsilon_{N}}{N\varepsilon_{N}^{+}(\eta)+\varepsilon_{N}^{2}(\eta)\varepsilon_{N}^{-}/\varepsilon_{N}^{2}}\ ,\nonumber \\
B'_{N} & = & \frac{2\sqrt{I^{0}C}}{I^{0}+C}B_{N}\ .\label{eq:Odd Bell}\end{eqnarray}
\textcolor{black}{Here $\varepsilon_{N}^{+}(\eta)=\varepsilon_{N}(\eta)+4$,
and $B_{N}$ is defined as in Eq (\ref{eq:Even Bell}).}

\textcolor{black}{}%
\begin{figure}[h]
\textcolor{black}{\includegraphics[width=1\columnwidth,height=4.5cm]{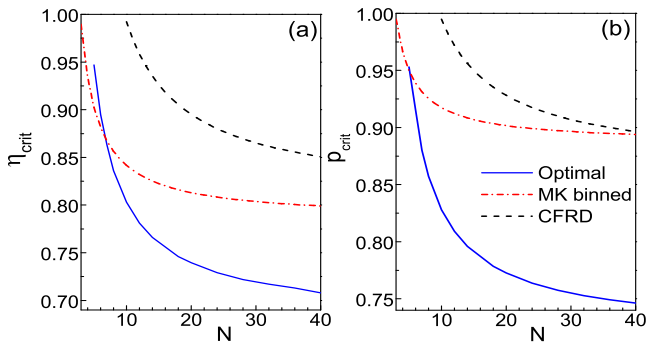}}

\textcolor{black}{\caption{\textcolor{black}{(a) The }critical\textcolor{black}{{} $\ $minimum
detection efficiency $\eta_{crit}$ for pure state, and (b) the critical
purity $p_{crit}$ for ideal detectors required for violation of functional
moment, CFRD} and MK inequalities with optimal choice of parameters.\label{fig:loss&noise}}
}
\end{figure}

\textcolor{black}{This approach is applied to enable a prediction
of the effect of loss and noise on the functional inequalities, and
the results are plotted in Fig. \ref{fig:loss&noise}. }These results
can be compared with the MK binning approach. With the choice of \textcolor{black}{optimal
angles}, we find the values of the MK Bell observable with binned
cv outcomes is \textcolor{black}{$B_{N}(p,\eta)=\frac{\sqrt{2}}{2}(\frac{4\eta p^{2}}{\pi})^{N/2}$,}
which gives the effect of detection inefficiencies and noise for the
optimal choice of angles. That implies a critical minimum efficiency
and purity $(\eta p^{2})_{crit}=2^{(1-2N)/N}\pi$ in order to violate
the inequality.\ \textcolor{black}{For lower $N$, the strategy of
binning and using the MK inequality shows an advantage, by allowing
a violation for $N=3,4,5$,---but even if $p=1,$ high efficiencies
$\eta>0.99$, $0.93$, $0.90$ are required. While high detection
efficiencies are feasible for homodyne detection, these efficiency
and purity values are still quite challenging once generation losses
are also taken into account. In view of this, the high requirement
for $\eta_{crit}$ for the case $N=3$ may be prohibitive. }

\textcolor{black}{These results show that the functional inequality
has much greater robustness against noise and inefficiency than the
MK inequality.} \textcolor{black}{For $N>7$, the functional cv inequality
used with an }\textit{\textcolor{black}{optimal}}\textcolor{black}{{}
function allows violation of LHV at much lower efficiencies and larger
maximum noise. The asymptotic decoherence product is $(p\eta)_{\infty}\sim0.6918$
in the large N limit. For a moderate efficiency $\eta_{crit}\sim80\%$
one requires $N=10$ if the optimized function, while the binned MK
case requires $N\sim40$.}

In conclusion, we have developed a new direction for the analysis
of cv nonlocality. For the input state treated here, the optimal measured
function always has the same functional form apart from changing the
parameter $\epsilon$, but more generally, the functional form may
depend on the experimental decoherence. Future research may include
further optimization of the functions for different entangled states
and application of this method to tests of other forms of nonlocality---i.e.,
entanglement \cite{entanglement criteria} and EPR steering \cite{Steering}.

\textcolor{black}{We wish to acknowledge funding for this project
from the Australian Research Council through a Discovery grant and
the ARC Centre of Excellence for Quantum-Atom Optics. EGC acknowledges
discussions with Antonio Acin, Daniel Cavalcanti and Yeong-Cherng
Liang.}


\begin{thebibliography}{16}
\bibitem{LHS}Bearing in mind these measurements can be noncompatible,
the LHS is simply an abbreviation for the expansion in terms of observable
moments, e.g. $\bigl\{\langle x_{1}x_{2}\rangle-\langle p_{1}p_{2}\rangle\bigl\}^{2}+\bigl\{\langle x_{1}p_{2}\rangle+\langle x_{2}p_{1}\rangle\bigl\}^{2}$
when $N=2$.
\end{thebibliography}
\end{document}